\documentclass[a4paper,11pt]{article}
\pdfoutput=1 

\usepackage{jcappub}
\usepackage{graphicx}
\usepackage{hyperref}
\usepackage{aas_macros}
\usepackage{amsmath}
\usepackage{amssymb}
\usepackage{xcolor}
\usepackage{slashed}
\usepackage{nccmath}
\usepackage{jcappub} 
\usepackage{lineno}
\usepackage{natbib}
\usepackage{ulem}
\usepackage[T1]{fontenc}
\usepackage[utf8]{inputenc}
\usepackage{aas_macros} 

\title{\boldmath Little Red Dots from Ultra-Strongly Self-Interacting Dark Matter}

\author[a,b]{M. Grant Roberts,}
\author[a]{Lila Braff,}
\author[a]{Aarna Garg,}
\author[a,b]{Stefano Profumo,}
\author[a,b]{and Tesla Jeltema}

\affiliation[a]{Department of Physics, University of California, Santa Cruz (UCSC),
Santa Cruz, CA 95064, USA}
\affiliation[b]{Santa Cruz Institute for Particle Physics (SCIPP),
Santa Cruz, CA 95064, USA}
\emailAdd{migrober@ucsc.edu}

\abstract{We investigate the possibility that the recently identified population of high-redshift, obscured quasars - known as ``Little Red Dots'' (LRDs) - originates from early black hole seed formation driven by ultra-strongly self-interacting dark matter (uSIDM). In this framework, dark matter halos undergo gravothermal core collapse due to large self-interaction cross sections, resulting in the rapid formation of massive black hole (BH) seeds with masses \( \gtrsim 10^{5}\,M_\odot \) at redshifts \( z \gtrsim 5 \). We develop a semi-analytic model that tracks the evolution of the dark matter halo population, the redshift of collapse \( z_{\rm coll} \), and the corresponding BH mass function. Black hole growth is modeled stochastically via a log-normal Eddington ratio distribution and a finite duty cycle.
We find that the uSIDM scenario naturally reproduces key observed properties of LRDs, including their abundance, compactness, and characteristic BH masses, while offering a mechanism for early, obscured black hole formation that is difficult to achieve in standard CDM-based models. The predicted SMBH mass function at \( z \sim 5 \) shows excellent agreement with LRD observational data and SIDM merger-tree simulations, particularly at the high-mass end (\( \mBH \gtrsim 10^{7}\,M_\odot \)). These results suggest that LRDs may serve as powerful observational tracers of exotic dark sector physics and that SMBH formation in the early universe could be significantly shaped by non-gravitational dark matter interactions.
}

\begin{document}

\newcommand{\mBH}{m_{\text{BH}}}
\newcommand{\mBHseed}{\mBH^{\text{seed}}}
\newcommand{\mBHobs}{\mBH^{\text{obs}}}
\newcommand{\mBHobsi}{m_{\text{BH},i}^{\text{obs}}}
\newcommand{\mBHtheory}{\mBH^{\text{theory}}}
\newcommand{\zcoll}{z_{\text{coll}}}
\newcommand{\zvir}{z_{\text{vir}}}
\newcommand{\zobs}{z_{\text{obs}}}
\newcommand{\cross}{\sigma/m}
\newcommand{\msun}{M_{\odot}}
\newcommand{\tsal}{t_{\text{sal}}}
\newcommand{\trel}{t_{\text{rel}}}
\newcommand{\rhocrit}{\rho_{\text{crit}}}
\newcommand{\cmg}{\text{cm}^{2}\text{g}^{-1}}
\newcommand{\kms}{\text{km}~\text{s}^{-1}}
\newcommand{\angstrom}{\r{A}}
\newcommand\sbullet[1][.5]{\mathbin{\vcenter{\hbox{\scalebox{#1}{$\bullet$}}}}}
\newcommand{\chisq}{\chi^{2}}
\newcommand{\Vmax}{V_{\text{max}}}

\newcommand{\lrb}[1]{\left[{#1}\right]}
\newcommand{\lrp}[1]{\left({#1}\right)}
\newcommand{\lrcb}[1]{\left\{{#1}\right\}}
\newcommand{\lrv}[1]{\left|{#1}\right|}
\newcommand{\lra}{\longrightarrow}

\newcommand{\M}{\mathcal{M}}
\newcommand{\lag}{\mathcal{L}}
\newcommand{\lagdark}{\lag_{\rm{dark}}}

\newcommand{\mchi}{m_{\chi}}
\newcommand{\mchione}{m_{\chi_1}}
\newcommand{\mchitwo}{m_{\chi_2}}
\newcommand{\mphi}{m_{\phi}}
\newcommand{\alphachi}{\alpha_{\chi}}
\newcommand{\alphachione}{\alpha_{\chi_{1}}}
\newcommand{\alphachitwo}{\alpha_{\chi_{2}}}
\newcommand{\gchione}{g_{\chi_{1}}}
\newcommand{\gchitwo}{g_{\chi_{2}}}
\newcommand{\ma}{m_{A'}}

\newcommand{\sigmam}{\sigma/m}
\newcommand{\sigmaeff}{\sigma_{\text{eff}}}
\newcommand{\sigmavel}{\sigma_\text{1D}}
\newcommand{\sigmaeffm}{\sigmaeff/m}

\newcommand{\chibar}{\overline{\chi}}
\newcommand{\OmegaDM}{\Omega_{\text{DM}}}
\newcommand{\OmegaCDM}{\Omega_{\text{CDM}}}
\newcommand{\OmegaDMh}{\OmegaDM\text{h}^{2}}
\newcommand{\OmegaCDMh}{\OmegaCDM\text{h}^{2}}

\newcommand{\alphaethos}{\alpha_{\rm{ETHOS}}}
\newcommand{\vrel}{v_{\rm{rel}}}
\newcommand{\kinmix}{\varepsilon_{\gamma}}
\newcommand{\fend}{f_{\rm uSIDM}^{\rm \infty}}

\newcommand{\spr}[1]{{\color{red}\bf[SP:  {#1}]}}
\newcommand{\grantcomment}[1]{{\color{blue}\bf[GR:  {#1}]}}

\newcommand{\refcomment}[1]{{\color{red}\bf[{#1}]}}

\newcommand{\tesla}[1]{{\color{cyan}\bf[TJ:  {#1}]}}

\maketitle
\flushbottom

\section{Introduction}

The discovery of ``Little Red Dots'' (LRDs) through deep infrared observations with the \textit{James Webb Space Telescope} (\textit{JWST}) has revolutionized our understanding of supermassive black hole (SMBH) formation and growth in the early Universe. These enigmatic objects represent a population of compact, heavily dust-reddened active galactic nuclei (AGN) observed at high redshifts ($z > 4$), characterized by their distinctive red rest-frame optical emission coupled with blue ultraviolet (UV) slopes \cite{Kokorev_2024,Labbe:2023ab}. The emergence of LRDs as a distinct class of high-redshift sources has fundamentally challenged existing models of early quasar evolution and SMBH assembly, necessitating a comprehensive reevaluation of the mechanisms driving black hole growth during the first billion years of cosmic history.

LRDs are believed to represent a critical evolutionary phase of obscured SMBHs in their earliest stages of quasar development, where vigorous accretion processes occur within heavily dust-enshrouded environments \cite{2025arXiv250604350Z,2025arXiv250305537I}.
These systems are thought to typically harbor supermassive black holes with masses spanning $10^6$ to $10^9$ $M_{\odot}$, undergoing episodes of rapid, often super-Eddington accretion that drive their exceptional luminosities despite significant obscuration \cite{Kocevski:2024zzz,Matthee:2024abc}; however, more observational data is needed to nail down the exact SMBH masses \cite{2025arXiv250509669S}. 

Recent deep X‑ray observations have cast serious doubt on super‑Eddington accretion as the origin of Little Red Dots (LRDs). Sacchi and Bogd\'an performed an ultra‑deep stack of Chandra exposures targeting 55 LRDs ($\sim$400 Ms total) and reported a robust non‑detection in the 0.3–7 keV band \cite{2025arXiv250509669S}. They argue that even mildly super‑Eddington accretion would unavoidably produce soft X‑rays above their sensitivity limit, unless obscured by a column density as extreme as \(N_H\gtrsim10^{25}\,\mathrm{cm}^{-2}\). Since JWST spectroscopy shows no compelling evidence for such extreme obscuration, the most parsimonious explanation is that the accretion is intrinsically sub‑Eddington, or that the black hole masses and bolometric luminosities have been overestimated.

Theoretical studies reinforce these observational constraints. Kido et al. \cite{Kido:2025} explore the consequences of embedding a super‑Eddington disc within a massive optically thick envelope to avoid disruptive outflows \cite{Kido:2025}. Although this envelope could stabilize the system, it is highly contrived and fails to naturally reproduce the observed spectral energy distributions. Meanwhile, alternative scenarios invoking sub‑Eddington fueling of primordial black hole seeds offer a more straightforward match to LRD abundance, luminosity, and red colors \cite{Huang:2024vwx}. In particular, the sub‑Eddington accretion onto supermassive primordial seeds naturally accounts for the X‑ray faintness, compact sizes, and moderate luminosities without requiring finely tuned obscuration or envelopes. Thus, both empirically and theoretically, super‑Eddington models are increasingly disfavored in favor of simpler, sub‑Eddington interpretations.

The demographic significance of LRDs becomes apparent when considering their remarkable abundance: they outnumber their unobscured UV-bright quasar counterparts by $2$--$3$ orders of magnitude at comparable redshifts, revealing a previously hidden population that may dominate the census of actively accreting black holes in the early Universe. This numerical dominance has profound implications for our understanding of the integrated black hole accretion history and suggests that the majority of early SMBH growth occurred in heavily obscured phases that were largely invisible to pre-\textit{JWST} observations.

The remarkable properties of LRDs - particularly their rapid SMBH growth rates and unexpectedly high abundance - have motivated the exploration of alternative formation scenarios that extend beyond conventional astrophysical mechanisms. One particularly intriguing possibility involves the role of exotic dark matter physics, specifically self-interacting dark matter (SIDM) and its potential connection to the observed LRD population. Unlike standard cold dark matter, SIDM allows dark matter particles to undergo significant collisional interactions, fundamentally altering the internal structure and evolution of dark matter halos \cite{Tulin:2017ara,Elbert:2014bma}. In sufficiently dense environments, these interactions can drive a phenomenon known as core collapse, where energy redistribution within the halo leads to a runaway increase in central density that could profoundly influence the formation and growth of embedded SMBHs.

The core collapse mechanism in SIDM halos presents a compelling framework for understanding the rapid assembly of massive black hole seeds in the early Universe \cite{fangzhou_LRD,2025arXiv250617641F,Feng2021}. During the collapse process, the dramatic concentration of dark matter in halo centers creates gravitational potential wells capable of efficiently funneling baryonic material into compact regions, potentially facilitating the formation of intermediate-mass black hole seeds with masses on the order of $10^4$--$10^6$ $M_{\odot}$ \cite{Huang:2024vwx}. These substantial seed masses could provide a significant head start for subsequent accretion processes, offering a natural explanation for how SMBHs with masses approaching $10^9$ $M_{\odot}$ could assemble within the limited timescales available in the first billion years after the Big Bang. This scenario addresses one of the most pressing challenges in early Universe SMBH formation: the difficulty of growing black holes from stellar-mass seeds to supermassive scales through standard accretion processes alone within the available cosmic time.

The SIDM framework also provides a potential explanation for several puzzling observational characteristics of LRDs. The high abundance of these systems at $z \sim 5 - 10$, where they outnumber unobscured quasars by orders of magnitude, could reflect the prevalence of SIDM core collapse events during epochs when dark matter densities and halo interaction rates were conducive to such phenomena. Furthermore, the remarkably compact nature of LRDs (typically $< 100$ pc) might be a natural consequence of baryon concentration within collapsed dark matter cores, where the enhanced gravitational focusing creates the dense, heavily obscured environments characteristic of these systems \cite{Zhang:2024pqr}. The extreme accretion rates inferred for LRD black holes could similarly emerge from the efficient matter transport enabled by the steep gravitational potential gradients present in core collapsed halos.

The remainder of this study is structured as follows: the following Section~\ref{sec:uSIDM} describes the theoretical framework of ultra-strongly self-interacting dark matter (uSIDM), including the physics of gravothermal core collapse and its implications for early black hole seed formation. We explore the relevant conditions for collapse, the associated timescales, and the dependence on halo mass and redshift. Section~\ref{sec:massfunction} presents predictions for the supermassive black hole (SMBH) mass function resulting from uSIDM-driven seed formation. This includes a Monte Carlo model incorporating halo growth, collapse redshifts, stochastic accretion histories, and the quasar duty cycle. We compare our results with both JWST-observed LRD mass functions and SIDM-based simulation outputs. Finally, we conclude by discussing observational prospects and potential avenues to test the uSIDM scenario using future multi-wavelength surveys and large-scale structure studies.

\section{Ultra-Self-Interacting Dark Matter and Early Galactic Core Collapse}\label{sec:uSIDM}

The Standard Model of Cosmology, while remarkably successful in explaining large-scale structure formation and cosmic evolution, faces significant challenges in accounting for the rapid assembly of supermassive black holes observed at high redshifts. The existence of quasars powered by $\sim 10^9$ $M_{\odot}$ black holes at $z > 6$, corresponding to cosmic ages of less than one billion years, represents one of the most pressing puzzles in modern astrophysics \cite{Banados:2018dla,Wang:2021qsv}. Traditional formation mechanisms, including Population III stellar remnants, direct collapse black holes, and hierarchical merging scenarios, struggle to produce such massive objects within the available time constraints, even under the most optimistic assumptions for accretion efficiency and duty cycles \cite{Volonteri:2010bc,Inayoshi:2020def}, though see \cite{Qin2025} for a new potential mechanism for direct collapse to occur. This tension has motivated the exploration of alternative formation pathways that invoke physics beyond the standard cold dark matter (CDM) paradigm, with ultra-self-interacting dark matter (uSIDM) emerging as a particularly promising candidate for facilitating the rapid formation of massive black hole seeds. In this section, we review the uSIDM framework, its potential for forming black hole seeds, and its potential implications for the origin of LRDs.

\subsection{Theoretical Framework of Ultra-Self-Interacting Dark Matter}

uSIDM represents an extension of the standard SIDM paradigm, characterized by significantly enhanced cross-sections, $\lrp{\cross}_{\rm{uSIDM}}~>> ~\lrp{\cross}_{\rm{SIDM}}$, for dark matter particle interactions that can reach values of $\sigma/m \gtrsim 10~\cmg$  in the velocity range relevant for galactic halos \cite{Tulin:2017ara,Kaplinghat:2016jlq}. 
uSIDM scenarios allow for velocity-dependent interaction cross-sections that can be dramatically enhanced at the low velocities characteristic of early, low-mass halos \cite{Loeb:2011aa,Feng:2009hw}. This velocity dependence is crucial for understanding the behavior of dark matter in the high-redshift Universe, where typical halo velocities were significantly lower than those observed in present-day galactic systems. For an explicit exploration of the uSIDM microphysics, see Ref.~\cite{Roberts_uSIDM_pheno}, but here we give a short overview of the particle physics, starting with the Lagrangian:

\begin{eqnarray}
 &&\lagdark = \overline{\chi}_{1}\lrp{i\slashed{\partial} - \mchione}\chi_{1} + \overline{\chi}_{2}\lrp{i\slashed{\partial} - \mchitwo}\chi_{2} - \gchione\chibar_{1}A'^{\mu}\gamma_{\mu}\chi_{1} - \gchitwo\chibar_{2}A'^{\mu}\gamma_{\mu}\chi_{2}\\
 && ~~~~~~~~~~~ -\frac{1}{4}F'^{\mu\nu}F'_{\mu\nu} - \frac{1}{2}m_{A'}^{2}A'^{\mu}A'_{\mu}\nonumber.
\label{eq:model-lagrangian}
\end{eqnarray}

Thus the model consists of three particles, one uSIDM particle $\chi_{1}$, one SIDM particle $\chi_{2}$, and the dark photon mediator of the self interactions $A^{\prime}$. Both $\chi_1$ and $\chi_2$ couple to $A^{\prime}$ in the same way, but the $\chi_1$ coupling, $\gchione$ is much larger than the $\chi_2$ coupling, $\gchitwo$, i.e., $\gchione >> \gchitwo$. To compensate for this hierarchy of couplings, the uSIDM particle makes up a small fraction, $f$, of the overall dark matter abundance. The remaining dark matter abundance, $\lrp{1-f}$, is comprised entirely of the SIDM particle. We have found that this framework is cosmologically consistent in terms of reproducing the correct dark matter relic density, $\OmegaDMh = \Omega_{\chi_1}h^{2} + \Omega_{\chi_2}h^{2} = 0.12$.

The enhanced interaction rates in uSIDM models fundamentally alter the thermal evolution and structural properties of dark matter halos, particularly during the early phases of cosmic structure formation. In conventional CDM scenarios, dark matter halos maintain relatively stable density profiles over cosmic time, with central densities that evolve primarily through accretion and merging processes. However, in uSIDM systems, the frequent collisional interactions between dark matter particles drive significant heat conduction within halos, leading to the formation of isothermal cores and the redistribution of kinetic energy from the halo outskirts toward the central regions \cite{Spergel:1999mh,Dave2001}. This process can initially expand halo cores, creating the characteristic flat, i.e. cored, density profiles that help resolve tensions with small-scale structure observations; see \cite{Oman:2015xda,deBlok:core_cusp,Tulin:2017ara,Adhikari:2022sbh}. However, as halos continue to accrete mass and their central densities increase, the enhanced interaction rates can eventually drive a phase transition known as core collapse or the gravothermal catastrophe \cite{Balberg:2002ya,Koda:2011aa,Outmezguine:gravothermal}.

The gravothermal instability in uSIDM halos represents a critical transition point where the balance between heat conduction and gravitational contraction breaks down, leading to a runaway collapse of the central dark matter core \cite{Balberg:2002ya,Koda:2011aa,Outmezguine:gravothermal}. This instability typically occurs when the central density of a self-interacting halo exceeds a critical threshold that depends on the interaction cross-section, halo mass, and velocity dispersion. For uSIDM models with large cross-sections, this threshold can be reached relatively early in cosmic history, particularly in the dense environments characteristic of high-redshift galaxy formation regions. The onset of core collapse is marked by a dramatic steepening of the central density profile, with the collapse proceeding on timescales much shorter than the typical dynamical times of the host halos.

\subsection{Environmental Conditions and Halo Mass Dependence}\label{sec:environmental-conditions}

The efficiency of uSIDM core collapse and its potential for producing quasar progenitors depends critically on the environmental conditions and halo properties present during the epoch of high-redshift galaxy formation. Our previous fits of uSIDM halos to high redshift quasars indicate that the onset of core collapse is most favorable in halos with masses in the range $10^{7.5}$--$10^{11.5}$ $M_{\odot}$, corresponding to the typical mass scale of galaxies hosting high-redshift quasars \cite{Roberts_uSIDM}. Below this mass range, halos lack sufficient gravitational binding energy to drive the collapse process to completion, while above this range, the increased velocity dispersions tend to suppress the interaction rates and delay the onset of instability. In either case, whether above or below the typical mass range, these halos will still contain some uSIDM particles - but will not undergo the efficient core collapse process. Instead, they will undergo the core collapse transition in a timescale roughly similar to the slower SIDM timescale. 

The redshift dependence of uSIDM core collapse also plays a crucial role in determining the epoch of massive black hole formation. At higher redshifts, the enhanced cosmic density and reduced halo velocity dispersions create conditions that are particularly favorable for driving gravothermal instabilities \cite{Elbert:2014bma,Koda:2011aa,Outmezguine:gravothermal}. Thus, the number of uSIDM produced high-redshift quasars is roughly fixed after the collapse redshift. This redshift dependence naturally explains the observed prevalence of high-redshift quasars and as the environmental conditions become less conducive to uSIDM core collapse in more recent cosmic epoch, this provides a physical mechanism for the apparent decline in quasar space density at lower redshifts (roughly fixed number, but increasing spacial volume).

The clustering properties and large-scale environment of uSIDM halos undergoing core collapse may also exhibit distinctive signatures that could be accessible to observational surveys. Halos experiencing gravothermal instabilities are expected to show enhanced central concentrations, modified mass-to-light ratios, and potentially altered correlation functions compared to their CDM counterparts \cite{Kamada:2017gzm,Correa:2021fkr}. These environmental signatures could provide important diagnostics for identifying quasar host galaxies that have undergone uSIDM core collapse, offering a pathway for observationally testing this formation mechanism.

\subsection{Observational Signatures and Constraints}

The uSIDM paradigm for high-redshift quasar formation makes several specific predictions that can be tested against current and future observational data. Perhaps most importantly, quasars formed through this mechanism should preferentially inhabit dark matter halos with distinct structural properties, including flattened central density profiles prior to collapse and potentially enhanced central concentrations following the instability \cite{Zavala:2019gpx,Sagunski:2021cji}. These structural differences could manifest in gravitational lensing observations, satellite kinematics, or the detailed dynamics of gas and stars within quasar host galaxies.

The formation timescales and efficiency of uSIDM core collapse also predict specific correlations between quasar properties and their large-scale environments. Quasars formed through this mechanism should show enhanced clustering on intermediate scales, reflecting the preferential formation of massive black holes in regions of enhanced dark matter density where core collapse is most efficient \cite{Turner:2021bos}. Additionally, the mass function of high-redshift quasars should exhibit characteristic features related to the halo mass dependence of the core collapse process, potentially providing a distinctive signature that could distinguish uSIDM formation from conventional astrophysical mechanisms.

\subsection{Implications for Little Red Dots and Early Quasar Evolution}

The uSIDM framework provides a compelling theoretical foundation for understanding the formation and properties of Little Red Dots as a distinct population of high-redshift quasars. The rapid assembly of massive black hole seeds through core collapse naturally explains the ability of LRDs to reach supermassive scales within the limited time available at $z > 4$, addressing one of the primary challenges posed by these systems \cite{Matthee:2024abc}. Moreover, the enhanced accretion rates enabled by the deep potential wells created during core collapse could account for the extreme luminosities and rapid variability observed in many LRD systems.

The compact sizes and high obscuration fractions characteristic of LRDs may also reflect the environmental conditions present during uSIDM core collapse. The efficient concentration of baryonic material into small spatial scales during the collapse process could naturally produce the dense, dusty environments that give LRDs their distinctive red optical colors and heavy extinction \cite{Wang:2024ghi,Kokorev_2024}. Additionally, the enhanced matter densities present in core collapsed halos could drive elevated star formation rates and dust production, contributing to the obscured nature of these systems.

The high abundance of LRDs compared to unobscured quasars at similar redshifts finds a natural explanation within the uSIDM framework, as the core collapse mechanism could be significantly more efficient at producing massive black holes than conventional formation pathways \cite{Kocevski:2024zzz,Pacucci:2024def}. The environmental dependence of core collapse also suggests that LRDs should exhibit distinctive clustering properties and preferentially inhabit regions of enhanced dark matter density, predictions that could be tested through large-scale structure studies and environmental analyses of LRD host galaxies.

\section{The uSIDM Model and the SMBH Mass Function}\label{sec:massfunction}

\subsection{The uSIDM Model}\label{subsec:uSIDM-model-equations}
The predictions we present for the SMBH mass function rely on the uSIDM seeding mechanism first developed in detail in our previous work~\cite{Roberts_uSIDM}. In this framework, a subdominant component of the dark matter - characterized by a large self-interaction cross section - undergoes gravothermal collapse within dark matter halos at high redshift. The onset of collapse is governed by the condition
\begin{equation}
    t(z_{\rm coll}) - t(z_{\rm vir}) = 455.65 \, t_{\rm rel}(f, \cross, m_{200}, \zvir, c_{200}),
\label{eq:solve_zcoll}
\end{equation}
where the relaxation time $t_{\rm rel}$ depends on the uSIDM fraction $f$, the cross section per unit mass $\sigma/m$, halo mass $m_{200}$, and halo concentration $c_{200}$ - for which we use Diemer19 \cite{Diemer_2019}. We model the velocity dependent cross-section as a simple power-law with $n=-4$ as this yields fits to high-z quasars from uSIDM and is the correct scaling relation of the uSIDM microphysics \cite{Roberts_uSIDM,Roberts_uSIDM_pheno}:

\begin{equation}
    \sigma(v)/m = \sigma/m \left(\frac{m_{200}}{10^{12}\ m_{\odot}}\right)^{n/3}.
\end{equation}

\noindent An explicit expression for $t_{\rm rel}$ is given by
\begin{equation}
    t_{\rm rel} = 0.354\,\mathrm{Myr}
    \left( \frac{m_{200}}{10^{12} M_\odot} \right)^{-1/3}
    \left( \frac{k_c(c_{200})}{k_c(9)} \right)^{3/2}
    \left( \frac{c_{200}}{9} \right)^{-7/2}
    \left( \frac{\rho_{\rm crit}(z)}{\rho_{\rm crit}(z=15)} \right)^{-7/6}
    \left( \frac{f \cross}{1\,\cmg} \right)^{-1},
\end{equation}
with $k_c(x) \equiv \ln(1+x) - x/(1+x)$ and $\rho_{\rm crit}(z)$ the critical density.

The black hole seed mass produced by this collapse process is then given by~\cite{Pollack:2015iug}
\begin{equation}
    \frac{\mBHseed(z)}{m_{200}} \simeq \frac{0.025 f}{\ln{(1 + c_{200})} - \frac{c_{200}}{1 + c_{200}}}.
\label{eq:mBH-calc}
\end{equation}

\noindent Once formed, the BH seeds grow exponentially via accretion until the redshift of observation:
\begin{equation}
    \mBH(\zobs) = \mBHseed(\zcoll)\exp({\lambda N_{e}}).
\label{eq:seed-growth}
\end{equation}
where the number of e-folds of growth $N_e$ is determined by the time elapsed and the Salpeter timescale $t_{\rm Sal}$ \cite{Salpeter} (assuming Eddington-limited accretion with radiative efficiency $\epsilon_r = 0.1$):
\begin{equation}
    t(z_{\rm obs}) = t(z_{\rm coll}) + N_e \, t_{\rm Sal}, \quad 
    t_{\rm Sal} \approx 45.1\,\mathrm{Myr}.
\label{eq:solve_Ne}
\end{equation}

Crucially, we emphasize that the entire collapse and seeding calculation depends only on the product $f \sigma/m$, so that the degeneracy between a low-$f$/high-$\sigma$ regime and a high-$f$/low-$\sigma$ regime is inherent to the model. As we showed in~\cite{Roberts_uSIDM}, this gives rise to a bimodal structure in viable parameter space, with distinct implications for halo environments and subsequent SMBH accretion histories.

The full derivation and a comprehensive statistical treatment of this model, including its application to JWST-observed quasars, is presented in~\cite{Roberts_uSIDM}. Readers are encouraged to consult that work for detailed equations of motion, MCMC constraints, and cross-section parameterizations across velocity-dependent scenarios.

\subsection{The SMBH Mass Function}

Understanding the SMBH mass function in the early universe provides a crucial test of models for BH seeding and growth. In this section, we explore how the uSIDM  framework predicts the mass function of BHs at high redshift. In particular, we compare our predictions to the implied black hole mass function based on LRD observations given observationally-motivated assumptions on accretion rates and active fraction.

We begin by modeling the growth of dark matter halos, which we assume to follow the form derived in Eq. 4 of ref.~\cite{halo_grow_paper}:

\begin{equation}
m_{200}(z_{i}) = m_{200}(z_{f})\lrp{1+\Delta z}^{\beta}e^{-\gamma \Delta z},
\label{eq:m200-growth}
\end{equation}

\noindent where $\beta = 0.1$ and $\gamma = 0.69$ provide a good fit across a broad range of halo masses, and $\Delta z = z_{i} - z_{f}$. Here, $z_i$ and $z_f$ denote the initial and final redshifts of halo evolution, respectively. This functional form enables us to evolve a present-day halo mass $m_{200}(z_f)$ backward in redshift to determine its progenitor mass $m_{200}(z_i)$ or vice versa.

To compute the black hole mass function (BHMF), we relate it to the halo mass function via a chain of Jacobians:

\begin{equation}
\frac{dN_{BH}}{d\mBH} = \frac{dN_{BH}}{dN_{H}} \frac{dN_{H}}{dm_{200}} \frac{dm_{200}}{d\mBH} = \frac{dN_{BH}}{dN_{H}} \frac{dN_{H}}{d\ln m_{200}} \frac{1}{m_{200}} \frac{dm_{200}}{d\mBH},
\label{eq:BHMF}
\end{equation}

\noindent where $H$ denotes halos and we assume a monotonic and deterministic mapping between halo mass and the central BH mass. Explicitly, $\frac{dN_{BH}}{dN_{H}}$ is the number of BHs per halo, $\frac{dN_{H}}{d\ln m_{200}}$ is the underlying halo mass function, and $ \frac{dm_{200}}{d\mBH}$ is the derivative of the halo mass to BH mass relation - which is determined by Eq.~\ref{eq:mBH-calc}. Changing the derivative to be with respect to $\log_{10} \mBH$, more suitable for plotting and interpretation, yields:

\begin{equation}
\frac{dN_{BH}}{d\log_{10}{\mBH}} = \ln{10}\frac{dN_{BH}}{dN_{H}} \frac{dN_{H}}{d\ln m_{200}} \frac{\mBH}{m_{200}} \frac{dm_{200}}{d\mBH}.
\label{eq:BHMF-lnMBH}
\end{equation}

This approach assumes a fixed BH accretion rate $\lambda$ for all objects. However, in reality, $\lambda$ is expected to vary stochastically across the BH population. In particular, LRD quasars are observed to follow a log-normal distribution in $\lambda$, with mean $\lambda_0 \approx 0.2$ and dispersion $\sigma_\lambda \approx 0.3$ \cite{fangzhou_LRD}, though the exact mean and dispersion can vary slightly depending on the redshift \cite{Xiao_2021,Willott_2010}. To model this, we treat the growth of each BH probabilistically via a joint distribution:

\begin{equation}
\frac{d^2N_{BH}^{\rm grow}}{d\log_{10}{\mBH}~d\lambda} = \varepsilon_{\rm DC}P(\lambda)\frac{dN_{BH}^{\rm seed}}{d\log_{10}{\mBH^{\rm seed}}},
\label{eq:2D-joint-distribution}
\end{equation}

\noindent where $P(\lambda)$ is the log-normal distribution governing accretion, and $\varepsilon_{\rm DC}$ is the quasar duty cycle (the fraction of black holes that are active at a given time). We take $\varepsilon_{\rm DC} = 0.01$ at $z \sim 5$, consistent with estimates in ref.~\cite{duty_cycle_z=5}.

The generation of the BH seed population, $\mBHseed$, at its collapse redshift $z = \zcoll$, requires care. $\zcoll$ depends not only on halo properties at virialization ($m_{200}(\zvir)$, $c_{200}(\zvir)$), but also on the uSIDM particle parameters, notably the uSIDM fraction, $f$, and the self-interaction cross-section, $\sigma/m$. Conceptually, we list the steps for the Monte Carlo as follows. To estimate $\zcoll$, we draw halos from the Tinker halo mass function \cite{Tinker2008} at redshift $z = \zvir$, thus we have $m_{200}(\zvir)$ and $c_{200}(\zvir)$, and then solve Eq.~\ref{eq:solve_zcoll} for $\zcoll$. We then evolve these halos masses forwards in time using Eq.~\ref{eq:m200-growth}, from $\zvir$ to $\zcoll$. Now that $\zcoll$ is known, we compute the BH seed masses at $z = \zcoll$ from Eq.~\ref{eq:mBH-calc}. To evolve the BH seeds forward under stochastic accretion via Monte Carlo sampling of Eq.~\ref{eq:2D-joint-distribution}, we first find the number of e-folds of accretion by solving Eq.~\ref{eq:solve_Ne} with $\zobs = 5$, then sample from the log-normal $\lambda$ distribution, and use Eq.~\ref{eq:seed-growth} to estimate the BH mass at $\zobs$, $\mBHobs$. We can then assemble the resulting BHMF at $z \sim 5$, corresponding to the LRD quasar epoch.

To be specific, the BHMF is estimated by binning in $\log_{10}m_{BH}$ over our total number of Monte Carlo samples, $N_{mc}$, with corresponding weights, $w_{i}$:

\begin{eqnarray}
    \frac{dN_{BH}}{d\log_{10}{\mBH}} = \sum_{i ~\in~\rm{bin}}^{N_{mc}}\frac{w_{i}}{\Delta(\log_{10}\mBH)} = \frac{1}{\Delta(\log_{10}\mBH)}\sum_{i ~\in~\rm{bin}}^{N_{mc}}w_{0}\varepsilon_{\rm DC}\,,
\end{eqnarray}

\noindent where $w_{i} = w_{0}\varepsilon_{\rm DC}$ and $\Delta(\log_{10}\mBH)$ is the bin width of our samples of $\log_{10}\mBH$. Each Monte Carlo sample corresponds to a comoving sub-population of BHs represented with weight $w_i$. The duty cycle multiplies the number density, not the mass, i.e. it is the fraction of BHs active at a given snapshot, not a rescaling of their intrinsic accretion rates. In particular, $w_{0} = \frac{n_{\rm seed}}{N_{mc}}$, such that $n_{\rm seed}$ is the original number density of seeds:

\begin{equation*}
    n_{\rm seed} = \int_{\log_{10}m_{200}^{\rm min}}^{\log_{10}m_{200}^{\rm max}} \frac{dN_{H}}{d\log_{10}m_{200}}d\log_{10}m_{200}~,
\end{equation*}

\noindent where $m_{200}^{\rm min} = 10^{7.5}~\msun$ and $m_{200}^{\rm max} = 10^{11.5}~\msun$ are the rough masses where gravothermal evolution is efficient in the early universe; see Section~\ref{sec:environmental-conditions}.

We now move to discuss the impact of mergers and luminosity observability cuts. For mergers, we consider a toy merger model. This is a simplified, analytic scenario with merger rates based on simulations. We assume a cut-off mass threshold, $m_{200}^{\text{cut}} \sim 10^{10.6}~\msun$, that determines which halos will be the primary halos, or secondary halos. In other words, if a given halo has $m_{200} < m_{200}^{\rm{cut}}$, then it is a secondary halo and might merge into one of the primary halos. We do not expect many mergers among more massive halos, as their merger rate is a sharply decreasing function of mass ratios. From the Millennium simulations~\cite{Millennium_sims}, we find that halos above this cut-off mass range will merge on average five times with smaller halos until $z\sim5$. So from our sample of primary halos, we take five of the secondary halos and merge them into the primary. However, to simulate some uncertainty in this average merger number, we only select 80\% of the primary halos to actually merge. Prior to binning, we implement a minimum luminosity cut, corresponding to roughly the smallest observed LRD luminosity, $L_{\text{min}}\sim4\times10^{43}~\rm{erg/s}$. The luminosity cut is defined by $\lambda L_{\text{Edd}} > L_{\text{min}}$; BH's that meet this criterion are selected into the final sample of BHs we consider for the mass function. Both of these effects contribute to the shape of the BHMF. Mergers increase the normalization, as well as flatten the lower end of the BHMF - whereas the luminosity cut turns the flattening downwards.

\subsection{Results}\label{subsec:results}

\begin{figure}[ht]
\includegraphics[scale=1]{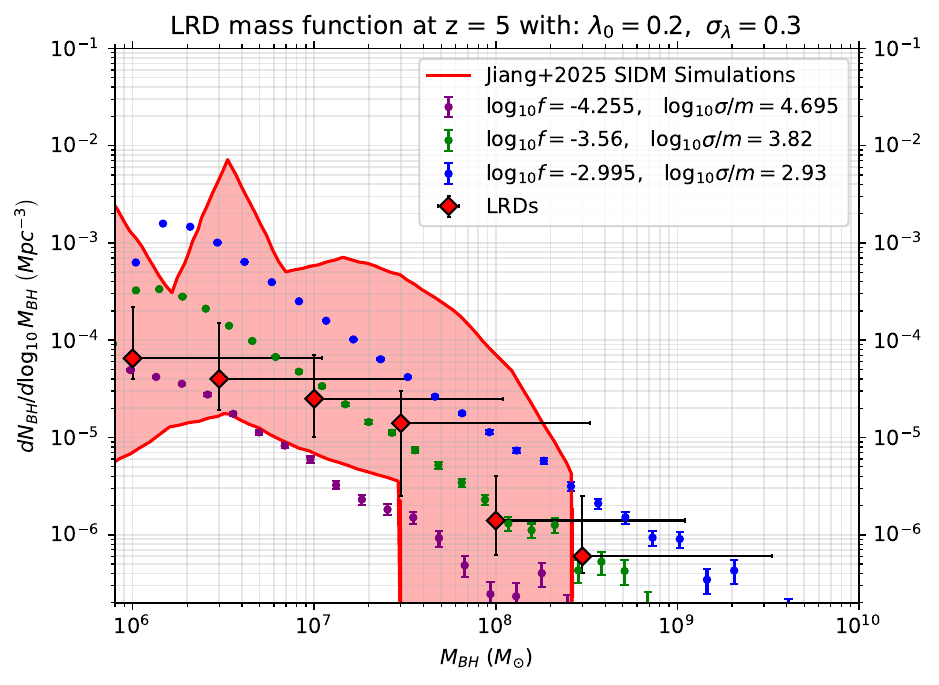}
\caption{We show the uSIDM LRD mass function (colored dots) for different uSIDM parameters taken from Fig. 3 in \cite{Roberts_uSIDM}; we also plot the Poisson error for our uSIDM LRD mass function in each mass bin. The uSIDM results are plotted against the SIDM simulation results from \cite{fangzhou_LRD} (shaded red), and the LRD mass function data points from \cite{Kokorev_2024} (red diamonds). The red diamonds are the inferred LRD black hole masses assuming Eddington accretion while the error bars show the range of black hole masses for accretion down to $\lambda=0.1$. We also note that the red shaded region from \cite{fangzhou_LRD} assumes all of the black holes in their simulations are active, while we assume a duty cycle of 1\% based on ref.~\cite{duty_cycle_z=5}.}
\label{fig:uSIDM-LRD-mass-function}
\end{figure}

In Fig.~\ref{fig:uSIDM-LRD-mass-function}, we compare our uSIDM-based predictions (colored points) with two reference datasets: the LRD mass function measurements from Kokorev et al.~\cite{Kokorev_2024} (red diamonds), and SIDM-based simulations from Jiang et al.~\cite{fangzhou_LRD} (red shaded region). The three uSIDM curves correspond to different combinations of cross-section and cooling parameters ($\log_{10} f$ and $\log_{10} \sigma/m$), taken from the parameter space explored in Roberts et al.~\cite{Roberts_uSIDM}. Specifically we use, $\lrp{\log_{10} f, ~\log_{10} \sigma/m} = \lrp{-4.255,~4.695},~\lrp{-3.56,~3.82},~\lrp{-2.995,~2.93}$, which are the purple, green, and blue dots respectively. We also plot the Monte Carlo Poisson error in each bin. Each uSIDM model successfully tracks the flattening of the BHMF at the lower mass end and the steepening of the BHMF at the higher mass end ($\mBH \lesssim 10^7 M_\odot$), where the SIDM model begins to underproduce BHs relative to LRD observations.  We also note that the SIDM simulations from Jiang et al.~\cite{fangzhou_LRD} assume that all of their BHs are active, so one should take the red shaded regions as an upper limit. We also mention that in ref.~\cite{fangzhou_LRD}, they do not plot the error bars for the LRD BH mass estimates with the data points they show - they assume  that the Eddington ratio is 1. The LRD points therefore represent a lower limit on the black hole masses, and in our Fig.~\ref{fig:uSIDM-LRD-mass-function}, we restore the error bars from ref.~\cite{Kokorev_2024}. We also mention that in the companion paper to \cite{fangzhou_LRD}, Shen et al. \cite{Shen2025}, they use a dissipative SIDM model to reproduce the SMBH and LRD mass functions. Similar to \citep{fangzhou_LRD}, they also require a duty cycle of 1 to match the LRD population (see their Figure 2).

The consistency between the uSIDM models and the observed LRD population at $z \sim 5$ provides a compelling case that strong self-interactions in the dark sector may play a role in early black hole formation, especially in producing the abundance of moderate-mass BHs required to match the data.

\subsection{Power-law Fitting Function}\label{subsec:power-law}

\begin{figure}[h]
\begin{minipage}{0.50\textwidth}
    \centering
    \includegraphics[width=\textwidth]{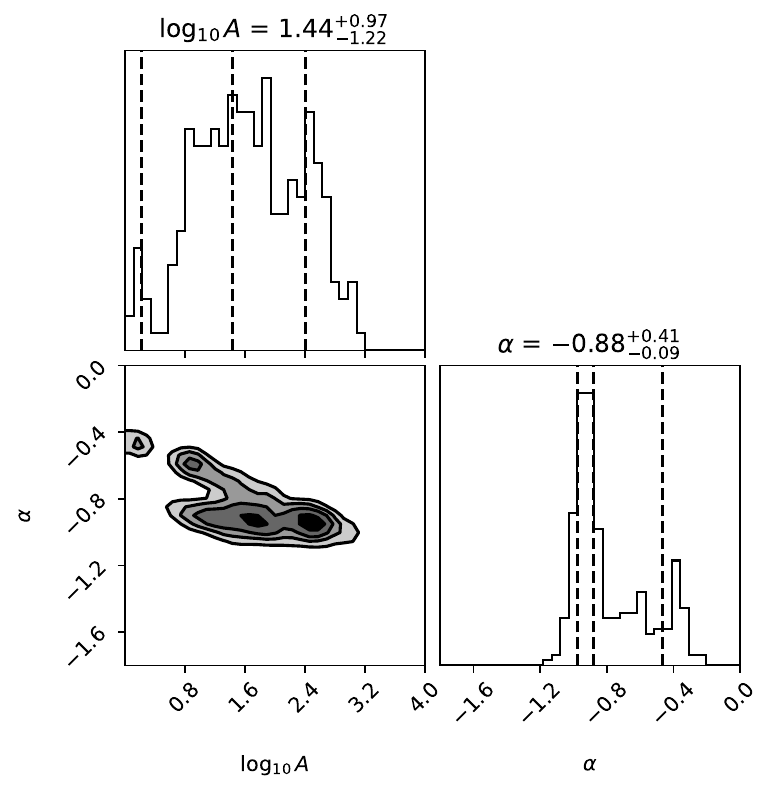}
\end{minipage}\hfill
\begin{minipage}{0.50\textwidth}
    \centering
    \includegraphics[width=\textwidth]{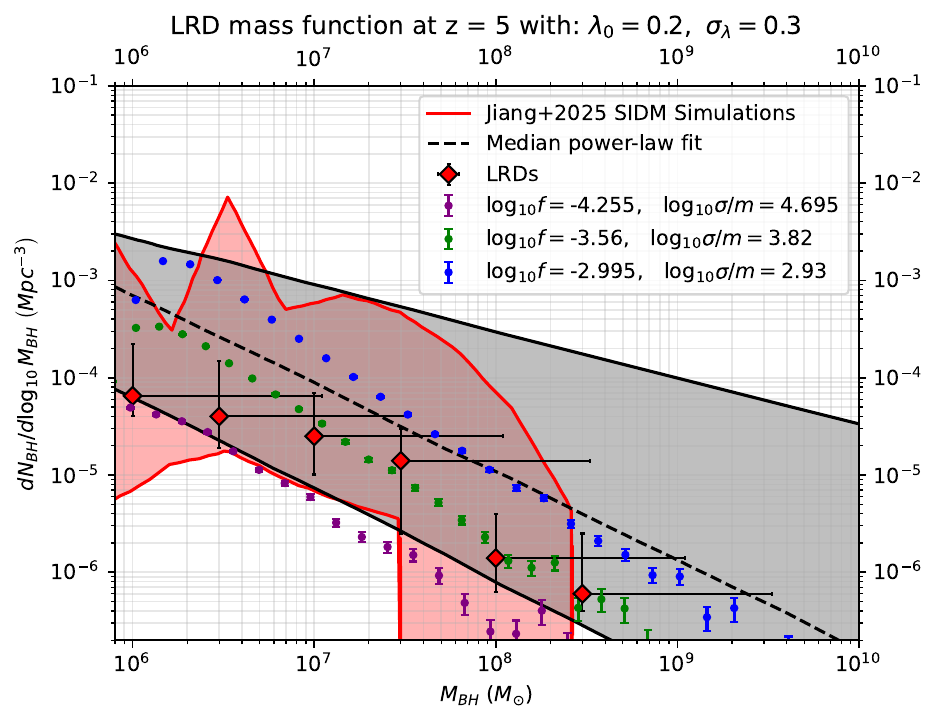}
\end{minipage}
\caption{We show the resulting distributions from fitting a power-law to 1000 realizations of the uSIDM parameter distribution from ref.~\cite{Roberts_uSIDM} (left). And we over plot the median and $\pm 1\sigma$ bands from the power-law realizations onto Fig.~\ref{fig:uSIDM-LRD-mass-function} (right).}
\label{fig:log10A-vs-alpha}
\end{figure}

In order to better characterize the uSIDM LRD mass function, we generate 1000 realizations of the uSIDM black hole mass function based on random draws of the uSIDM parameters $f$ and $\sigma/m$ from the 2D probability distribution from ref.~\cite{Roberts_uSIDM}. We then fit each of these uSIDM LRD mass function realizations, in the mass range of the LRD data points, with a power-law of the form:

\begin{equation}
    \frac{dN_{BH}}{d\log_{10}{\mBH}} = A~\mBH^{\alpha},
\label{eq:power-law-fit}
\end{equation}

\noindent where $A$ is a normalization parameter that in principle depends in a complicated way on the uSIDM parameters, and $\alpha$ is the log-slope. During the fitting, we also remove any uSIDM LRD mass function realizations that do not have a normalization within a few order of magnitudes of the LRD data. Similarly, we remove any realizations whose functional shape is a complete mismatch for the LRD data - such as a high peaking curve that falls off before covering the entire data range. The acceptable uSIDM parameters were approximately in the ranges $-5.53\lesssim~\log_{10} f~ \lesssim -1.59$ and $2.91 \lesssim~\log_{10}\cross~\lesssim4.73$ such that $0.5~\cmg~\lesssim f\cross~\lesssim10~\cmg$, the last is enforced by approximate optical thickness constraints \cite{Pollack:2015iug}.
In Fig.~\ref{fig:log10A-vs-alpha} (left) we plot the realizations that produce physically relevant mass functions. Because of the structure of Eq.~\ref{eq:BHMF-lnMBH}, we are inherently in the long power-law fall off of the halo mass function, thus we expect a power-law to fit the uSIDM LRD mass function well. We find that the uSIDM LRD mass function is well fit by a power-law with a median slope of $\alpha\sim-0.88$. In Fig.~\ref{fig:log10A-vs-alpha} (right), we plot the median and $1\sigma$ error bands of the power-law fit, onto the LRD mass function data and we can see that the realization fit does quite well in matching the data.

\section{Discussion}\label{sec:discussion}

Observational signatures of SIDM-driven LRD formation might manifest in several distinctive ways that could be accessible to current and future surveys. LRDs originating from SIDM core collapse could exhibit anomalous clustering properties or preferentially inhabit dark matter halos with unusual mass-to-light ratios compared to their standard cold dark matter counterparts. Environmental studies of LRD host halos might reveal overdensities or correlation functions that deviate from expectations based on conventional structure formation models. Additionally, the detailed kinematics of gas and stellar components within LRD host galaxies could potentially preserve signatures of the enhanced gravitational fields present during the core collapse phase, though disentangling such effects from standard AGN feedback processes remains a significant observational challenge.

However, connecting the observed properties of LRDs to underlying SIDM physics requires careful consideration of the complex interplay between dark matter dynamics and baryonic processes. While the weak or undetected X-ray emission characteristic of many LRDs \cite{Yue:2024jkl,Pacucci:2024def} could potentially reflect the extreme obscuration expected in core collapsed environments, current observations primarily attribute such properties to AGN-driven dusty outflows and associated feedback mechanisms \cite{Wang:2024ghi}. The challenge lies in identifying observational diagnostics that can definitively distinguish between SIDM-influenced formation scenarios and purely baryonic processes, as both could potentially produce systems with similar electromagnetic signatures.

Future observational strategies will be critical for advancing our comprehensive understanding of LRDs and testing alternative formation mechanisms, including potential connections to exotic dark matter physics. Planned studies leveraging \textit{JWST}'s unprecedented spectroscopic capabilities, combined with extended baseline variability monitoring and higher spatial resolution observations from ALMA, promise to address many of the outstanding questions surrounding these enigmatic objects. Complementary investigations focused on the large-scale environment and clustering properties of LRDs, potentially accessible through wide-field surveys and gravitational lensing studies, could provide crucial tests of SIDM-driven formation scenarios. These multi-faceted investigations will be essential for constraining the physical mechanisms driving LRD properties, establishing their evolutionary connections to other AGN populations, and ultimately determining whether exotic dark matter physics plays a role in supermassive black hole assembly in the early cosmos.

\section{Conclusion}\label{sec:conclusion}

The emergence of Little Red Dots as a dominant population of high-redshift, dust-obscured quasars has posed a serious challenge to conventional SMBH formation models. In this work, we have explored the hypothesis that these objects originate from black hole seeds formed via gravothermal core collapse in uSIDM halos. This process, which relies on large velocity-dependent self-interaction cross sections, enables the rapid concentration of dark and baryonic matter in early halos, producing BH seeds with masses \( \mBHseed \sim 10^{5} - 10^{7}\,M_\odot \) on timescales \( \lesssim 10^{8} \) years.

We presented a detailed semi-analytic framework for modeling the collapse redshift \( z_{\rm coll} \), halo mass evolution, and the subsequent black hole growth through stochastic Eddington-limited accretion. Our model incorporates a log-normal distribution of accretion rates and a finite quasar duty cycle \( \varepsilon_{\rm DC} \sim 0.01 \), in line with recent constraints at \( z \sim 5 \). Using Monte Carlo realizations, we computed the SMBH mass function at high redshift and compared it against observational data from JWST and simulation results from SIDM-based merger trees.

Our findings indicate that uSIDM models are able to reproduce the observed number density and mass distribution of LRDs without requiring fine-tuned or non-standard astrophysical assumptions. In particular, the steep rise in the low-mass end of the LRD mass function (\( \mBH \lesssim 10^{7}\,M_\odot \)) is captured naturally within the uSIDM framework, while CDM or vanilla SIDM scenarios underproduce this population, if they produce any LRDs at all. This suggests that LRDs could be the first empirical evidence of non-gravitational interactions in the dark matter sector operating at galactic scales during the early universe. In this vein, with forthcoming observational data of LRDs at different redshifts, we can track the evolution of the uSIDM LRD mass function's normalization and shape to further probe the uSIDM model. We necessarily expect these to change with redshift as the model has built in redshift dependence via the halo mass function and the merger rates from the Millennium Simulations.

Beyond mass functions, the uSIDM hypothesis makes several testable predictions. LRD host galaxies should preferentially reside in high-density environments conducive to early core collapse, exhibit compact stellar and gas distributions (on scales \( \lesssim 100\,\mathrm{pc} \)), and show signs of enhanced baryon inflow consistent with deep potential wells. Future observational programs - particularly those combining JWST spectroscopy, ALMA kinematic mapping, and large-scale structure clustering - could detect these signatures and discriminate between uSIDM and standard seeding pathways.

In summary, uSIDM offers a compelling explanation for the rapid emergence and abundance of obscured SMBHs in the early universe. If validated by further observations, this would mark a major breakthrough in our understanding of both black hole cosmology and the particle nature of dark matter. In future work, we will investigate more realistic merger scenarios from N-body simulations.

\acknowledgments
This work is partly supported by the U.S.\ Department of Energy grant number de-sc0010107 (SP). This work is based in part on observations made with the NASA/ESA/CSA James Webb Space Telescope. The data were obtained from the Mikulski Archive for Space Telescopes at the Space Telescope Science Institute, which is operated by the Association of Universities for Research in Astronomy, Inc., under NASA contract NAS 5-03127 for JWST.

\appendix

\newpage

\bibliographystyle{JHEP}
\bibliography{references}
\end{document}